\documentclass[12pt,reqno]{amsart}
\usepackage{graphicx,amsmath,amsfonts,amssymb,bbm}
\usepackage{psfrag}
\usepackage{epsf}
\usepackage{latexsym}
\usepackage{rotating}

\newtheorem{remark}{Remark}

\begin{document}

\title[Gaussian solitary waves and compactons in FPU]{
Gaussian solitary waves and compactons
in Fermi-Pasta-Ulam lattices with Hertzian potentials}
\author{Guillaume James$^{a}$ and Dmitry Pelinovsky$^{b}$}
\address{$^a$ INRIA Grenoble Rh\^one-Alpes and Laboratoire Jean Kuntzmann, Universit\'e de Grenoble and CNRS,
BP 53, Grenoble Cedex 9, France, 38041 \\
$^b$ Department of Mathematics, McMaster University,
Hamilton, Ontario, Canada, L8S 4K1}
\email{Guillaume.James@imag.fr (G.James)\\
dmpeli@math.mcmaster.ca (D.Pelinovsky)}

\date{\today}


\maketitle

\begin{abstract}
We consider a class of fully-nonlinear Fermi-Pasta-Ulam (FPU) lattices,
consisting of a chain of particles coupled by fractional power nonlinearities
of order $\alpha >1$. This class of systems
incorporates a classical Hertzian model describing acoustic wave propagation
in chains of touching beads in the absence of
precompression. We analyze the propagation of localized waves
when $\alpha$ is close to unity. Solutions varying
slowly in space and time are searched with an appropriate scaling,
and two asymptotic models of the chain of particles are derived consistently.
The first one is a logarithmic KdV equation, and possesses linearly orbitally stable
Gaussian solitary wave solutions. The second model consists of
a generalized KdV equation with H\"older-continuous
fractional power nonlinearity and admits compacton solutions, i.e.
solitary waves with compact support. When $\alpha \rightarrow 1^+$, we numerically establish
the asymptotically Gaussian shape of exact FPU solitary waves with
near-sonic speed, and analytically check the
pointwise convergence of compactons towards the limiting Gaussian profile.
\end{abstract}

\vspace{2ex}

\section{\label{intro}Introduction}

The problem of analyzing the response of a nonlinear lattice to a localized disturbance arises in many
applications, such as the study of stress waves in granular media after an impact \cite{neste2,sb},
the excitation of nonlinear oscillations in crystals by atom bombardment \cite{dcer,dub},
or the response of nonlinear transmission lines to a voltage pulse \cite{afshari}.
Several important dynamical phenomena can be captured by the Fermi-Pasta-Ulam (FPU)
model \cite{cam} consisting of a chain of particles coupled by a pairwise interaction potential $V$.
The dynamical equations for a spatially homogeneous FPU chain read
\begin{equation}
\label{nc}
\frac{d^2 x_{n}}{dt^2} =
V^\prime(x_{n+1}-x_n)-V^\prime(x_{n}-x_{n-1}),
\ \ \
n\in \mathbb{Z},
\end{equation}
where $x_{n}(t)\in \mathbb{R}$ is
the displacement of the $n$th particle from a reference position.
System (\ref{nc}) can be rewritten in terms of the relative displacements $u_{n}=x_{n}-x_{n-1}$
and particle velocities $p_n = \dot{x}_n$ as follows
\begin{equation}
\label{ncrelham}
\frac{d u_{n}}{dt}=p_{n}-p_{n-1}, \ \ \
\frac{d p_{n}}{dt} = V^\prime(u_{n+1})-V^\prime(u_{n}), \ \ \
n\in \mathbb{Z} .
\end{equation}
The dynamical evolution of localized solutions of (\ref{ncrelham})
is strongly influenced by the properties of the interaction potential $V$.
In its most general form, the interaction potential satisfies
\begin{equation}
\label{potential-V}
V \in C^2 (\mathbb{R}) \quad \mbox{\rm with} \;\;
V^\prime (x)=\kappa \, x + \mathcal{O}(|x|^\alpha ),
\end{equation}
where $\alpha >1$ and $\kappa \geq 0$.

In the work \cite{Mielke}, the dispersive stability of the zero equilibrium state
is proved for $\kappa >0$ and $\alpha > 4$, i.e. for sufficiently weak nonlinearities near the origin.
More precisely, the amplitude (i.e. supremum norm) of the solution of
the FPU lattice (\ref{ncrelham}) goes to $0$ when $t\rightarrow + \infty$
for all initial conditions sufficiently small in $\ell^1$, where
$\ell^1$ denotes the classical Banach space of bi-infinite summable sequences.

In contrast, in many situations nonlinear effects are strong enough to compensate
dispersion, yielding the existence of coherent localized solutions of
the FPU lattice (\ref{ncrelham}) such as solitary waves propagating at constant speed, or
time-periodic breathers (see e.g. \cite{cam} for a review).
The first existence result for solitary waves in
a general class of FPU lattices was obtained by Friesecke and Wattis \cite{friesecke},
when $V$ has a local minimum (not necessarily strict) at the origin
and is superquadratic at one side
(see also \cite{herrmann} and references therein). In addition,
the existence of solitary waves near the so-called
long wave limit was established in \cite{pego,iooss} for smooth ($C^3$) potentials $V$.
More precisely, for $\kappa>0$ and $V^{'''}(0)\neq 0$ (i.e. $\alpha =2$ in (\ref{potential-V})),
there exists a family of small amplitude solitary waves parameterized by their velocity
$c \gtrsim c_{\rm{s}} := \sqrt{\kappa}$, where $c_{\rm s}$ defines
the ``sound velocity" of linear waves. These solutions take the form
$$
u_n(t)=\kappa \, (4 V^{'''}(0))^{-1}\, \epsilon^2 \, z(\epsilon\, (n-c\, t))+ \mbox{higher order terms},
$$
where $\epsilon = \sqrt{24(c-c_{\rm{s}})/c_{\rm{s}}}$ and
$z(\eta ) =  \rm{sech}^2 (\eta /2)$.
In particular, these solitary waves decay exponentially in space and
broaden in the limit of vanishing amplitude. Equivalently, one has
\begin{equation}
\label{approxkdv}
u_n(t)=\kappa \, (4 V^{'''}(0))^{-1}\, \epsilon^2 \, y(\xi,\tau) + \mbox{higher order terms},
\end{equation}
where $\xi := \epsilon\, (n-c_{\rm{s}}\, t)$, $\tau := \epsilon^3\, c_{\rm{s}}\, t/24$, and
$y(\xi ,\tau ) := z(\xi -\tau )$ is a solitary wave solution of the Korteweg--de Vries (KdV) equation
\begin{equation}
\label{kdv}
\partial_\tau y + 3 y \, \partial_\xi y + \partial_\xi^3 y =0.
\end{equation}
More generally, the solutions $y\, : \, \mathbb{R} \times [0,T] \rightarrow \mathbb{R}$ of
the KdV equation (\ref{kdv}) yield solutions of the FPU system of the form (\ref{approxkdv}), valid on
a time scale of order $\epsilon^{-3}$ \cite{bp,kal,sw}.
In addition, the nonlinear stability of small amplitude FPU solitary waves was proved in
\cite{pego,hoffman3,mizu}, as well as the existence and stability of asymptotic
$N$-soliton solutions \cite{hoffman2,mizu3}.
These results allow to describe in particular the propagation of
compression solitary waves in homogeneous granular chains under precompression \cite{neste2}.

Another interesting case corresponds to fully-nonlinear interaction potentials, where
$\kappa =0$ (which corresponds to a vanishing sound velocity, that is, $c_{\rm s} = 0$)
and $V$ has a local minimum at the origin. A classical example is given by the
Hertzian potential
\begin{equation}
\label{vhertz}
V(x)=\frac{1}{1+\alpha}\, |x|^{1+\alpha}\,H(-x)
\end{equation}
with $\alpha >1$, where we denote by $H$ the Heaviside step function. This potential
describes the contact force between two initially tangent elastic bodies
(in the absence of precompression)
after a small relative displacement $x$ \cite{johnsonbook}.
The most classical case is obtained for
$\alpha = 3/2$ and corresponds to contact between spheres,
or more generally two smooth non-conforming surfaces.
More recently, granular chains involving
different orders of nonlinearity have
attracted much attention, see \cite{sekimoto,sunsen} and references therein.
In particular, experimental and numerical studies on solitary wave propagation
have been performed with chains of hollow spherical particles
of different width \cite{daraio} and chains of cylinders \cite{khatri},
leading to different values $\alpha$ in the range
$1.15 \leq \alpha \leq 1.5$
(see also \cite{sun} for other systems with $\alpha$ close to unity).

The propagation of stationary compression pulses in the FPU lattice (\ref{nc})
with potential (\ref{vhertz}) for $\alpha = 3/2$ was first analyzed
by Nesterenko \cite{neste2}. These results rely on a formal continuum
limit and provide approximate solitary wave solutions with compact
support. An alternate continuum limit problem
has been introduced in \cite{ap} for arbitrary values of $\alpha >1$,
leading to different (compactly supported) approximations
of solitary waves. The existence of exact solitary wave solutions of
the FPU lattice (\ref{nc}) with potential (\ref{vhertz}) follows from the general result of
Friesecke and Wattis \cite{friesecke} mentioned previously
(see also \cite{mackay,stef}). The width of these solitary waves is
independent of their amplitude due to the homogeneous nonlinearity of
the Hertzian potential. In addition, the
fully-nonlinear character of the Hertzian potential induces a
doubly-exponential spatial decay of solitary waves \cite{english,stef}.

While the above analytical results provide useful informations on
strongly localized solitary waves, they are not entirely satisfactory for
several reasons. First of all, the existence result of \cite{friesecke}
does not provide an approximation of the solitary wave profile,
and the approximations available in the literature \cite{ap,neste2}
rely on a ``long wave" assumption that is not justified
(for example, the solitary waves considered in \cite{neste2}
are approximately localized on five particles).
In addition, the dynamical properties of solitary waves in
fully-nonlinear FPU lattices are not yet understood.
Indeed, no mathematical results are available concerning their stability,
the way they are affected by lattice inhomogeneities, or
the existence of $N$-soliton solutions.
Another interesting problem is to characterize the excitation of one or several solitary waves from
a localized initial perturbation \cite{hinch,job}.
For $c_{\rm{s}} \neq 0$ and small amplitude long waves, this problem can be partially
analyzed in the framework of KdV approximation by using the inverse scattering transform methods
\cite{schuur}, but such reduction is presently unavailable for fully-nonlinear FPU lattices.
These questions are important for the analysis of
impact propagation in granular media, and more generally
for the design of multiple impact laws in multibody mechanical systems \cite{hinch,sb}.

In this paper, we attack the problem by considering a suitable long wave limit
of fully-nonlinear FPU lattices.
We consider the FPU lattice (\ref{nc}) with the
homogeneous fully-nonlinear interaction potential
\begin{equation}
\label{vhom}
V(x)=\frac{1}{1+\alpha}\, |x|^{1+\alpha},
\end{equation}
with $\alpha >1$.
Obviously, all solutions $u_n \leq 0$ of the FPU lattice (\ref{nc}) with the potential
(\ref{vhom}) are also solutions of the Hertzian FPU lattice (\ref{nc}) and (\ref{vhertz}).
The problem can be rewritten in terms of the relative displacements in the following way
\begin{equation}
\label{ncrel}
\frac{d^2 u}{dt^2} = \Delta \left( u\, |u|^{\alpha -1}\right),
\end{equation}
where we denote $u=(u_n )_{n\in \mathbb{Z}}$ and
$(\Delta\, u )_n=u_{n+1}-2\, u_{n}+u_{n-1}$ is the discrete Laplacian.
For approximating the temporal dynamics of (\ref{ncrel}) in a continuum limit,
fully-nonlinear versions of
the Boussinesq equation considered in \cite{ap,neste2} possess serious drawbacks,
since they may lead to blow-up phenomena
in analogy with the classical ``bad" Boussinesq equation \cite{yang}.
In section \ref{bouss}, we numerically show that these models
introduce artificial dynamical instabilities with arbitrarily large growth rates, which
suggests ill-posedness of these equations \cite{Wright2}.
Instead of using a Boussinesq-type model, we then
formally derive a logarithmic KdV (log-KdV) equation as a modulation
equation for long waves in fully-nonlinear FPU lattices, obtained in the limit
$\alpha \rightarrow 1^+$ (section \ref{deriv}). The log-KdV equation takes the form
\begin{equation}
\label{kdv-1}
\partial_\tau v + \partial^3_\xi v + \partial_\xi (v \ln{|v|} )=0
\end{equation}
and provides approximate solutions $u_n(t) \approx v(\xi , \tau)$
of the original FPU lattice (\ref{ncrel}) for $\xi := 2\sqrt{3}\, \epsilon (n-t)$, $\tau := \sqrt{3}\, \epsilon^3\, t $,
and $\epsilon := \sqrt{\alpha -1} \approx 0$.

The log-KdV equation (\ref{kdv-1}) admits Gaussian solitary wave solutions
(section \ref{trav}), which have been previously identified
as solutions of the stationary logarithmic nonlinear Schr\"odinger equation (log-NLS)
in the context of nonlinear wave mechanics \cite{bm}.
Closer to our case, Gaussian homoclinic solutions have been also found to approximate
the envelope of stationary breather solutions in Newton's cradle
(i.e. system (\ref{nc}) and (\ref{vhertz}) with an additional on-site potential)
in the limit $\alpha \rightarrow 1^+$ \cite{js12}.
In section \ref{trav}, we numerically check that solitary wave solutions of the Hertzian FPU lattice
with velocity $v_{\rm{s}}=1+c\, (\alpha -1)$
converge towards Gaussian approximations when $c\in \mathbb{R}$ is fixed and
$\alpha \rightarrow 1^+$. These solitary waves have velocities close to unity, which corresponds
to the value of sound velocity in the linear chain with $\alpha =1$.
In addition, we check that the FPU solitary waves are well approximated by
the compacton solutions derived in \cite{ap} when $\alpha \in (1,\frac{3}{2}]$.
To go beyond the stationary regime, we check numerically that the Gaussian approximation
captures the asymptotic shape of a stable pulse forming after a localized velocity perturbation
in the Hertzian FPU lattice (\ref{nc})-(\ref{vhertz}) with $\alpha \approx 1$
(section \ref{gauss}). Consistently with the above dynamical simulations,
we prove in section \ref{lstab}
the linear orbital stability of Gaussian solitary waves for the log-KdV equation.
Our analysis makes use of a suitable convex conserved
Lyapunov function, but
negative index techniques developed in recent works \cite{kapstef,dmitrystab}
for KdV-type equations would also apply.

The link between Gaussian solitary waves and compactons is made explicit is
section \ref{comprev}, where we check the pointwise convergence of the
compacton solutions of \cite{ap}
towards Gaussian profiles when $\alpha \rightarrow 1^+$.
In addition, following the methodology developed in section \ref{deriv}, we derive
from the fully-nonlinear FPU lattice a generalized KdV equation with H\"older-continuous nonlinearity (H-KdV):
\begin{equation}
\label{kdv-2}
\partial_\tau v + \partial^3_\xi v + \frac{\alpha}{\alpha -1}\, \partial_\xi ( v-v\, |v|^{\frac{1}{\alpha} -1} )=0 .
\end{equation}
When $\alpha \rightarrow 1^+$,
the H-KdV equation (\ref{kdv-2}) is consistent with the FPU lattice
in the sense that each solution to this equation
``almost" satisfies (\ref{ncrel}) up to a small residual error.
Equation (\ref{kdv-2}) admits explicit compacton solutions
whose form is close to the compactons
obtained in \cite{ap} with the use of a Boussinesq--type model.
When $\alpha \rightarrow 1^+$, these solutions converge towards
the Gaussian solitary waves studied in section \ref{trav}, and thus they
provide an (asymptotically exact)
approximation of FPU solitary waves with near-sonic speed.
This result sheds a new light on the
compacton approximations for FPU solitary waves heuristically derived
in the literature \cite{ap,neste2}.
Another interest of the H-KdV equation lies in the
(non-differentiable) H\"older-continuous nonlinearity $v\, |v|^{\frac{1}{\alpha} -1}$ which
allows for the existence of compactons. This type of degeneracy is
quite different from the classical feature of
compacton equations which incorporate degenerate nonlinear dispersion \cite{Wright2,rh}.

We finish this paper with a summary of our results and a
discussion of several open questions
concerning the qualitative dynamics of the log-KdV and H-KdV equations and their
connections with fully-nonlinear FPU chains (section \ref{alt}).

\section{\label{bouss}Fully nonlinear Boussinesq equation and compactons}

Fully nonlinear Boussinesq equations have been introduced in \cite{neste2,ap}
as formal continuum limits of FPU chains with Hertzian-type potentials.
In \cite{neste2}, the continuum limit is performed on system (\ref{nc})
with potential (\ref{vhom}) describing particle displacements, whereas \cite{ap} considers system
(\ref{ncrel}) for relative displacements. In what follows, we discuss the continuum
limit introduced in \cite{ap}, which takes a slightly simpler form than the system
derived in \cite{neste2}.

The fully nonlinear Boussinesq equation introduced in \cite{ap} takes the form
\begin{equation}
\label{B}
u_{tt} = (u\, |u|^{\alpha -1})_{xx} + \frac{1}{12} (u\, |u|^{\alpha -1})_{xxxx},
\end{equation}
where $u_{|x=n}$ denotes an approximation of a solution $u_n$ of (\ref{ncrel}).
The right side of (\ref{B}) is obtained by keeping the first two terms of the
formal Taylor expansion of the discrete Laplacian in (\ref{ncrel})
$$
\Delta = 2(\cosh{\partial_x}-1)=\sum_{k=1}^{+\infty}{\frac{2}{(2k) ! }\, \partial_x^{2k}}.
$$
This truncation is purely formal, but a numerical justification is presented
in \cite{ap} in the particular case of solitary wave solutions. More precisely,
the solitary waves $u_n(t) = w(z)$, $z = n-t$ of (\ref{ncrel})
are numerically compared with
solitary wave solutions $u(x,t)=w(z)$, $z=x-t$ of (\ref{B}).
For this class of solutions, equation (\ref{B}) reduces to a fourth order ordinary differential equation, which
can be integrated twice and leads to
\begin{equation}
\label{compacton-eq}
w = w\, |w|^{\alpha -1}  + \frac{1}{12} \frac{d^2}{d z^2}  w\, |w|^{\alpha -1},
\ \ \
z\in \mathbb{R},
\end{equation}
whereas equation (\ref{ncrel}) reduces to the differential advance-delay equation
\begin{equation}
\label{diffadvdel-eq}
\frac{d^2 w}{d z^2} = \Delta (w\, |w|^{\alpha -1}),
\ \ \
z\in \mathbb{R},
\end{equation}
with $(\Delta w )(z)=w(z+1)-2 w(z)+w(z-1)$.
The wave velocity can be normalized to unity due to a scaling invariance of
the FPU system (\ref{nc}) with homogeneous potential (\ref{vhom})
(or Hertzian potential (\ref{vhertz})), namely each solution $u_n$ generates
a one-parameter family of solutions
${| v_{\rm{s}} |}^{\frac{2}{\alpha -1}}\, u_n(v_{\rm{s}} t)$
with $v_{\rm{s}} \in \mathbb{R}$ (the same scaling invariance exists in system (\ref{B})).

According to the numerical computations presented in \cite{ap}, the solitary wave
of the differential advance-delay equation (\ref{diffadvdel-eq})
is well approximated by the compactly supported solitary wave
of the differential equation (\ref{compacton-eq}) for $\alpha =3/2$, and
the discrepancy increases with $\alpha$.
The compacton solution of (\ref{compacton-eq}) found in \cite{ap}
takes the form
\begin{equation}
\label{compacton}
w_{\rm{c}}(z) = \left\{ \begin{array}{ll} A \cos^{\frac{2}{\alpha-1}}(B z), & |z| \leq \frac{\pi}{2 B}, \\
0, & |z| \geq \frac{\pi}{2 B}, \end{array} \right.
\end{equation}
where
$$
A = \left( \frac{1+\alpha}{2 \alpha} \right)^{\frac{1}{1-\alpha}}, \quad B = \frac{\sqrt{3}(\alpha-1)}{\alpha}.
$$

In what follows, we reexamine the consistency of (\ref{ncrel})
and (\ref{B}) from a dynamical point of view, by analyzing the spectral stability of compactons.
Linearizing (\ref{B})
at the compacton $w_{\rm{c}}$ in the reference frame travelling with
unit velocity, we use the ansatz $u(x,t) = w_{\rm{c}}(x-t) + U(x-t) e^{\lambda t}$, where $\lambda$ is
the spectral parameter and $U$ is the perturbation term.
We arrive at the spectral problem
\begin{equation}
\label{B-lin}
{\left(\lambda - \frac{d}{dz} \right)}^2 U = \left( \frac{d^2}{dz^2} + \frac{1}{12}\,  \frac{d^4}{dz^4} \right) \left( k_\alpha\, U \right),
\end{equation}
where $k_\alpha (z) := \alpha w_{\rm{c}}^{\alpha - 1}(z)=\alpha A^{\alpha -1}\, \cos^2(B\, z) \,  \mathbbm{1}_{[-\frac{\pi}{2B},\frac{\pi}{2B}]}(z)$
and $\mathbbm{1}$ denotes the characteristic function. One can notice that $k_\alpha$ and $k_\alpha^{'}$ vanish at the end points
$z = \pm \frac{\pi}{2B}$ of the compact support of $w_{\rm{c}}$.
We look for eigenvectors in the Hilbert space
$$
D = \left\{ U \in H^2(\mathbb{R}), \quad k_\alpha\, U \in H^4(\mathbb{R}) \right\} .
$$
Since $k_\alpha^{''}$ and $k_\alpha^{'''}$ are discontinuous at $z = \pm \frac{\pi}{2B}$, this yields the condition
\begin{equation}
\label{B-bc}
U\left(\pm \frac{\pi}{2B}\right) = 0, \quad U'\left(\pm \frac{\pi}{2B}\right) = 0 .
\end{equation}
This allows us to reduce the eigenvalue problem (\ref{B-lin}) to
the compact interval $\left[-\frac{\pi}{2B},\frac{\pi}{2B}\right]$ with boundary
conditions (\ref{B-bc}), and approximate the spectrum
with the standard finite difference method
(we have used second-order difference approximations
for derivatives and $2000$ grid points).
If there exist
eigenvalues with ${\rm Re}(\lambda) > 0$, then the solitary wave is spectrally unstable.
If all the eigenvalues are located at the imaginary axis, then the solitary wave is called
spectrally stable.

\begin{figure}[h]
\begin{center}
\includegraphics[scale=0.3]{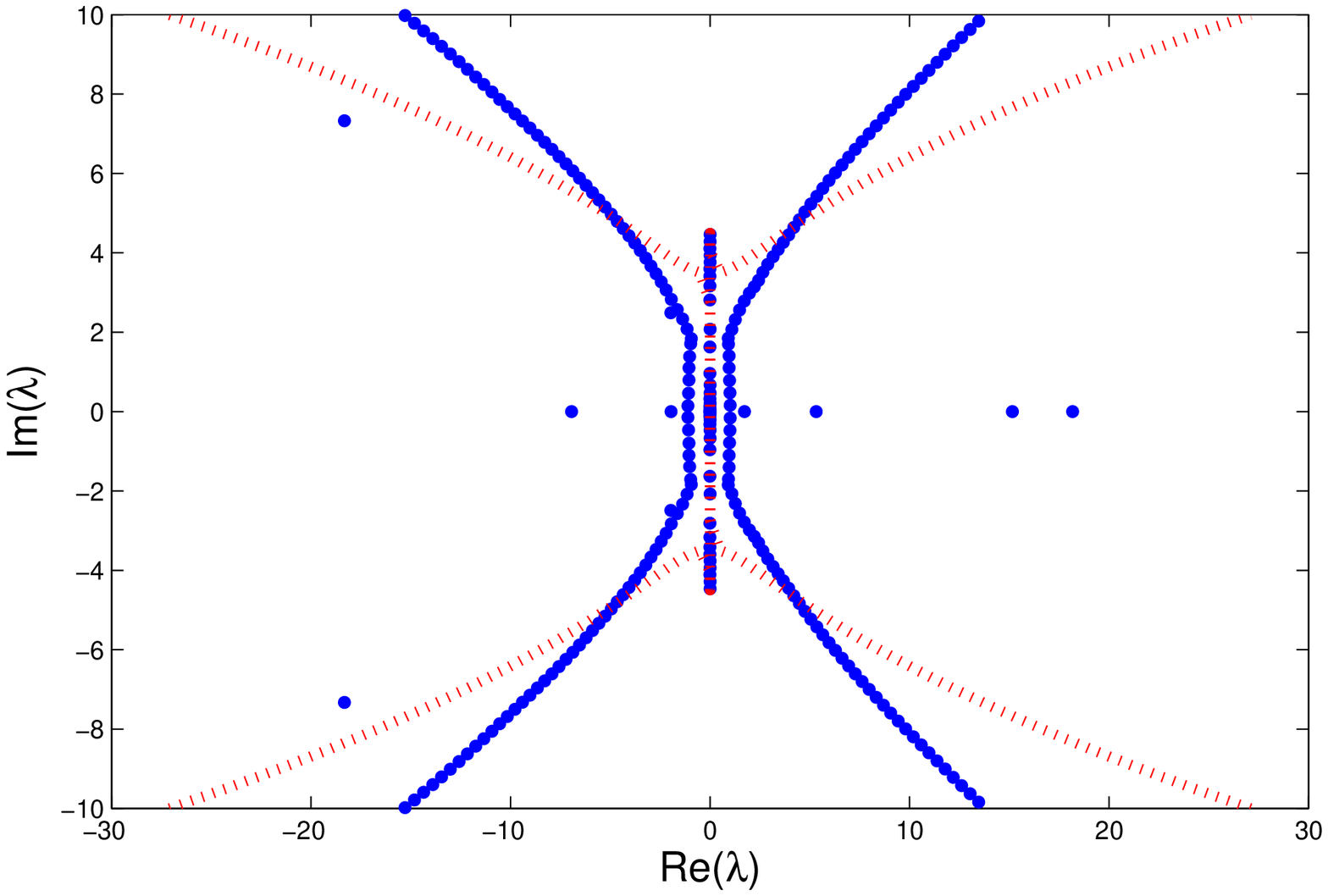}
\includegraphics[scale=0.3]{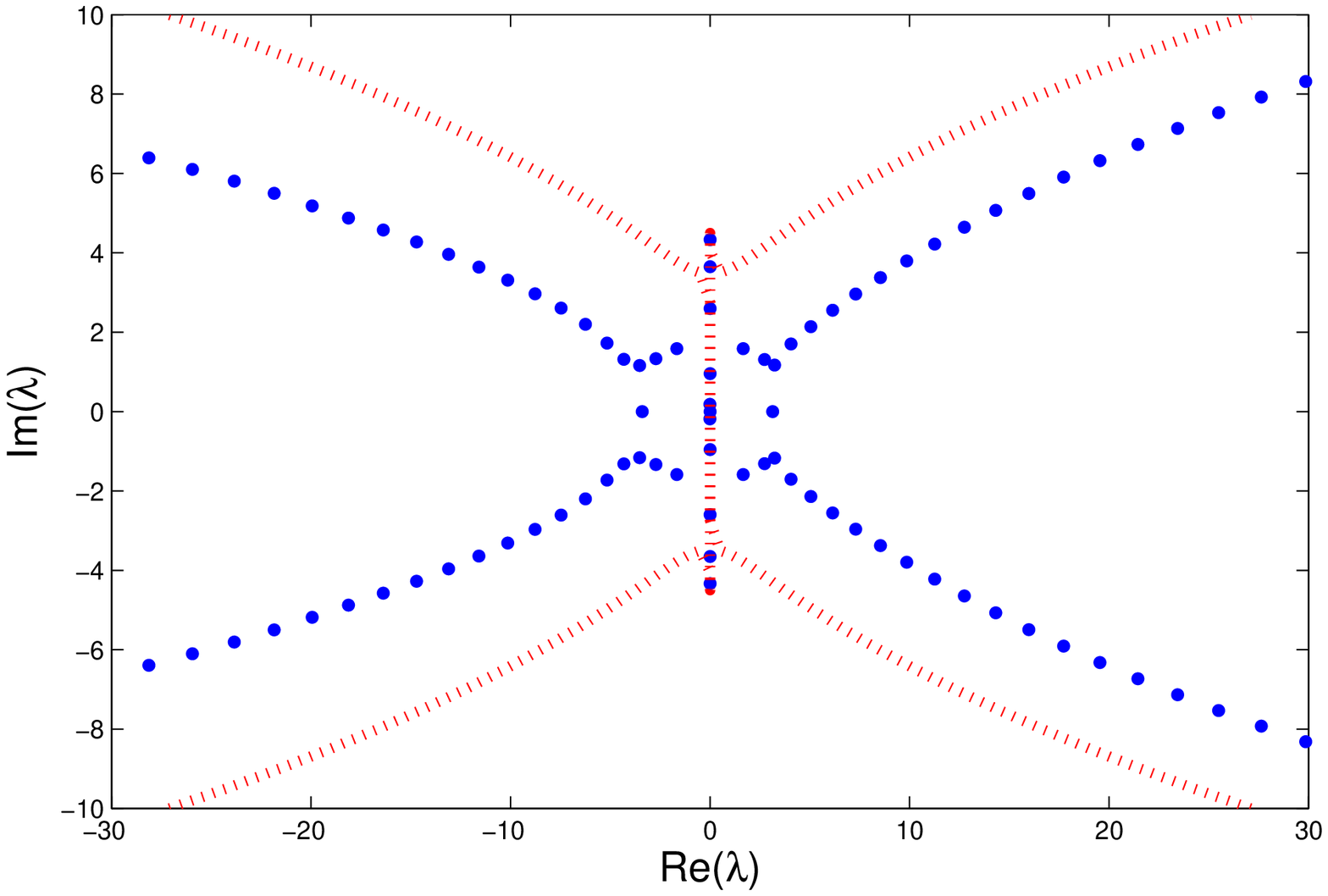}
\includegraphics[scale=0.3]{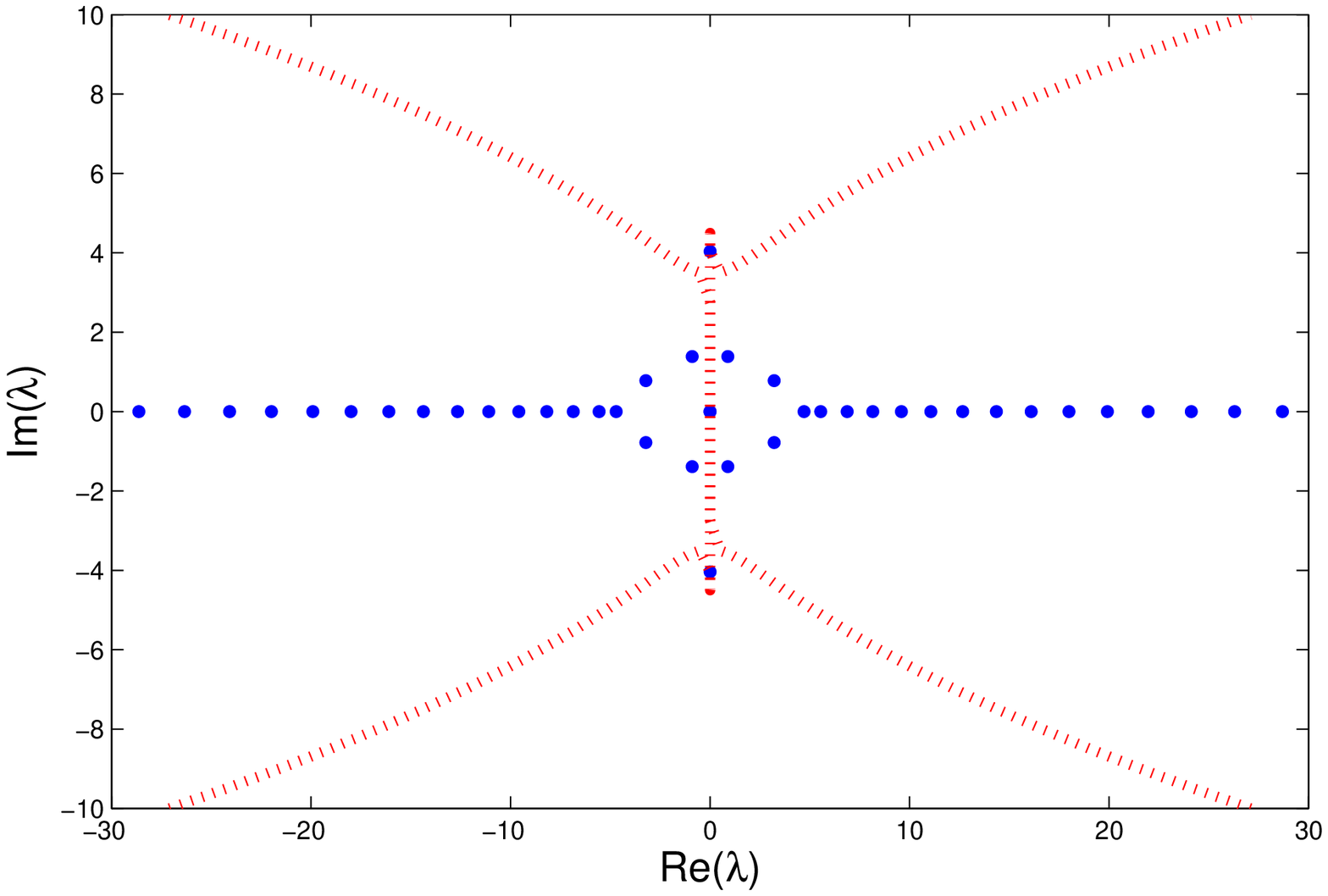}
\end{center}
\caption{\label{fig-spectrum}
Eigenvalues of the spectral problem (\ref{B-lin})-(\ref{B-bc})
(blue dots) for $\alpha = 1.05$ (top left), $\alpha = 1.2$ (top right), and $\alpha = 1.5$ (bottom).
The red dotted curves show the continuous spectrum (\ref{cont-spectrum})
obtained in the limit case $\alpha \rightarrow 1^+$.
}
\end{figure}

Figure \ref{fig-spectrum} shows
the complex eigenvalues $\lambda$ of the spectral problem
(\ref{B-lin})-(\ref{B-bc}) for $\alpha = 1.05, 1.2, 1.5$.
The spectrum is invariant under $\lambda \rightarrow -\lambda$
and $\lambda \rightarrow \bar\lambda$ (note the presence of a small number non-symmetric
eigenvalues, which originate from numerical errors).
We find the existence of unstable eigenvalues for all values of $\alpha >1$ considered, and
the eigenvalues approach the real line far from the origin
(this part of the spectrum is not visible in the first two panels of figure \ref{fig-spectrum}).
Consequently, these results
imply the spectral instability of the compacton (\ref{compacton}) in system (\ref{B}).
Note that the
usual notion of instability may not be well defined, since the evolution problem (\ref{B}) may
not be well posed. Indeed,
our numerical results indicate that the spectrum of (\ref{B-lin})-(\ref{B-bc})
is unbounded in the positive half-plane
(in fact at both sides of the imaginary axis), and thus the linearized
evolution problem may be ill-posed. We conjecture that ill-posedness occurs also in
system (\ref{B}), in analogy with ill-posedness results recently obtained
in \cite{Wright2} for certain nonlinear degenerate dispersive equations.

Along these lines, it is interesting to consider the limit case of
the spectral problem (\ref{B-lin})-(\ref{B-bc})
when $\alpha \rightarrow 1^+$. Since
$k_\alpha (z) \rightarrow 1$ for all $z \in \mathbb{R}$ and $B \rightarrow 0$,
the limiting spectral problem possesses constant coefficients and is defined on
the entire real line. Using the Fourier transform,  one can compute the
(purely continuous) spectrum explicitly, which yields
\begin{equation}
\label{cont-spectrum}
\lambda = \lambda_{\pm}(k) := i k \pm k \sqrt{\frac{1}{12} k^2 - 1}, \quad k \in \mathbb{R}.
\end{equation}
This limit case is represented in figure \ref{fig-spectrum} (red curves).
The spectrum being unbounded in the positive half-plane, the corresponding linear evolution problem
is then ill-posed.

The above instability phenomena are not physically meaningful since the solitary waves are known to be stable
from simulations of impacts in Hertzian chains \cite{neste2}.
In equation (\ref{cont-spectrum}) obtained in the limit $\alpha\rightarrow 1^+$, these instabilities
occur for short wavelengths (with $k^2 > 12$), whose dynamical evolution cannot be correctly captured by
the continuum limit (\ref{B}). In the next section,
we derive a different asymptotic model free of such artificial instabilities.

\begin{remark}
A classical model introduced to correct
artificial short wavelength instabilities of (\ref{B}) corresponds to
the regularized Boussinesq equation
\begin{equation}
\label{rb}
u_{tt} = (u\, |u|^{\alpha -1})_{xx} + \frac{1}{12} u_{xxtt}
\end{equation}
(see e.g. \cite{rosenau}).
This model has the inconvenience of
altering the spatial decay of solitary waves
in the case of fully-nonlinear interaction potentials.
Indeed, looking for traveling wave solutions $u(x,t)=w(z)$, $z=x-t$
and integrating equation (\ref{rb}) twice, one obtains
\begin{equation}
\label{rbs}
\frac{1}{12}\, \frac{d^2 w}{dz^2}=w - w\, |w|^{\alpha-1}
\end{equation}
after setting two integration constants to $0$.
Equation (\ref{rbs}) admits nontrivial symmetric
homoclinic solutions $\pm w_\alpha (z)$ satisfying
$\lim_{z \rightarrow \pm \infty}w_\alpha (z)=0$, corresponding
to solitary wave solutions of (\ref{rb}). These solutions decay exponentially
in space, 
which is too slow compared with the superexponential decay of the solitary
wave solutions of (\ref{diffadvdel-eq}).
\end{remark}

\section{\label{kdvloga}The log-KdV equation and Gaussian solitary waves}

\subsection{\label{deriv}Formal derivation of the log-KdV equation}

In order to pass to the limit $\alpha \rightarrow 1^+$ for long waves,
it is convenient to rewrite (\ref{ncrel}) in the form
\begin{equation}
\label{ncrel2}
\left( \frac{d^2}{dt^2} - \Delta \right)\, u= \Delta\, f_\alpha (u),
\end{equation}
where
\begin{equation}
\label{deffa}
f_\alpha (u) := u\, (|u|^{\alpha -1}-1)
= (\alpha -1)\, u \ln{|u|} + \mathcal{O}((\alpha -1)^2) 
\end{equation}
(uniformly in $u$ on bounded intervals)
when $\alpha \rightarrow 1^+$.
For $\alpha=1$, we have
$f_1(u) = 0$ for all $u \in \mathbb{R}$ and
system (\ref{ncrel2}) reduces to a semi-discrete linear wave
equation. In that case, the scaling in (\ref{approxkdv})
(with $c_{\rm{s}} =1$) yields a linearized KdV equation for the
envelope function $y$. To analyze the limit $\alpha \rightarrow 1^+$,
we assume the same type of scaling for the solution $u$, i.e. we
search for solutions depending on
slow variables $\xi := 2\sqrt{3}\, \epsilon (n-t)$
and $\tau := \sqrt{3}\, \epsilon^3\, t $, where $\epsilon > 0$ is a small parameter.
We look for solutions of the form
\begin{equation}
\label{scal}
u(t)= v(\xi , \tau) + \mathcal{O}(\epsilon).
\end{equation}
In contrast with (\ref{approxkdv}), the leading term $v(\xi,\tau)$ is assumed
of order unity and the remainder term in (\ref{scal}) is $\mathcal{O}(\epsilon)$.
Of course, due to the scaling invariance of the FPU lattice
(\ref{ncrel}) for $\alpha >1$,
solutions with arbitrarily small or large amplitudes can be deduced from
any solution of the form (\ref{scal}).

From the scaling (\ref{scal}) and using a Taylor expansion and the chain rule, we obtain
\begin{equation}
\label{oper1}
\Delta = 12\, \epsilon^2 \partial^2_\xi (1+ \epsilon^2 \partial^2_\xi + \mathcal{O}(\epsilon^4))
\end{equation}
and
\begin{equation}
\label{oper2}
\frac{d^2}{dt^2} - \Delta = -12\, \epsilon^4\, \partial_\xi (\partial_\tau + \partial^3_\xi )+\mathcal{O}(\epsilon^6).
\end{equation}
To evaluate the right side of (\ref{ncrel2}), we use
the expansion
$$
f_\alpha (u) = f_\alpha ( v + \mathcal{O}(\epsilon)) = (\alpha -1)\, v \ln{|v|} +
\mathcal{O}((\alpha -1)^2+|\alpha -1| \, \epsilon\,  |\ln{\epsilon}|) ,
$$
where the logarithmic remainder term accounts for the possible vanishing of $v$.
Setting now $\epsilon := \sqrt{ \alpha -1 } $ and using (\ref{oper1}), we obtain
\begin{equation}
\label{nl}
\Delta\, f_\alpha (u) = 12\, \epsilon^4\, \partial^2_\xi (v \ln{|v|} )+ \mathcal{O}(\epsilon^5 |\ln{\epsilon}|) .
\end{equation}
With this choice of $\epsilon$, the left- and right-hand sides of equation
(\ref{ncrel2}) have the same order $\epsilon^4$ according to expansions (\ref{oper2}) and (\ref{nl}).
Substituting these expansions in (\ref{ncrel2}) yields
\begin{equation}
\label{consistency}
\epsilon^4\, \partial_\xi \big(\partial_\tau v + \partial^3_\xi v + \partial_\xi (v \ln{|v|} )\big)+
\mathcal{O}(\epsilon^5 |\ln{\epsilon}|)=0 .
\end{equation}
Then neglecting the higher order terms and integrating with respect to $\xi$ leads to
\begin{equation}
\label{logkdv}
\partial_\tau v + \partial^3_\xi v + \partial_\xi (v \ln{|v|} )=0,  \ \ \
\xi \in \mathbb{R} ,
\end{equation}
where the integration constant has been fixed to $0$ in order to cover the case when
$\lim_{\xi \rightarrow +\infty}{v(\xi , \tau )}=0$. We shall call equation (\ref{logkdv}) the
logarithmic KdV (log-KdV) equation. It can be rewritten
\begin{equation}
\label{logkdvpot}
\partial_\tau v + \partial^3_\xi v + \partial_\xi (W^\prime (v) )=0,  \ \ \
\xi \in \mathbb{R} ,
\end{equation}
where the potential $W$ reads
$$
W (v)= \frac{v^2}{2} \big( \ln{|v|} - \frac{1}{2}\big).
$$
Equation (\ref{consistency}) shows that the log-KdV equation is consistent with
the nonlinear lattice (\ref{ncrel}), i.e. each solution of (\ref{logkdv}) is
almost a solution of (\ref{ncrel}) up to a small residual error.

Note that if $v$ is a solution of (\ref{logkdv}), so is $-v$.
In addition, equation (\ref{logkdv}) admits a nonstandard Galilean symmetry
involving a rescaling of amplitude, i.e.
each solution $v$ generates a one-parameter family of solutions
\begin{equation}
\label{scaling}
[\phi(c)\cdot v ] (\xi , \tau) := e^c\, v(\xi - c\, \tau , \tau), \quad c\in \mathbb{R}.
\end{equation}
In particular, all travelling wave solutions of the log-KdV equation
(\ref{logkdv}) can be deduced from its stationary solutions.
This property is inherited from the scaling invariance of
the FPU system (\ref{ncrel})
(this point will be detailed in section \ref{trav}).

Equation (\ref{logkdv}) falls within the class of generalized KdV equations.
Systems in this class possess three (formally) conserved quantities \cite{zhid}, namely
the mass
\begin{equation}
\label{mass}
M(v)  = \int_{\mathbb{R}}{v\, d\xi},
\end{equation}
the momentum
\begin{equation}
\label{momentum}
P(v) = \frac{1}{2} \int_{\mathbb{R}}{v^2\, d\xi},
\end{equation}
and the energy
\begin{equation}
\label{energy}
E(v) = \frac{1}{2}\int_{\mathbb{R}}{(\partial_\xi v)^2 \,d\xi} - \int_{\mathbb{R}}{W(v)\, d\xi}.
\end{equation}
Well-posedness results for the Cauchy problem associated with (\ref{logkdvpot})
are known when $W^{\prime\prime}$ is a $C^2$ function
\cite{zhid}, but the existing theory does not apply to
our case where $W^{\prime\prime} (v)=1+\ln{|v|}$ diverges logarithmically at the origin.

\subsection{\label{trav}Stationary solutions}

Looking for solutions of the log-KdV equation (\ref{logkdv}) depending only of $\xi$,
one obtains the stationary log-KdV equation
\begin{equation}
\label{slogkdv}
\frac{d^3 v}{d\xi^3} +  \frac{d}{d\xi}W^\prime (v)=0.
\end{equation}
Integrating once under the assumption $\lim_{\xi \rightarrow +\infty}{v(\xi)}=0$,
one obtains
\begin{equation}
\label{slognls}
\frac{d^2 v}{d\xi^2} + W^\prime (v) =0, \quad \Rightarrow \quad
\frac{d^2 v}{d\xi^2} + v \ln{|v|} =0 .
\end{equation}
This equation can be seen as a
(one-dimensional) stationary log-NLS equation \cite{bm}.

\begin{figure}[h]
\psfrag{u}[1][Bl]{ $v$}
\psfrag{W}[1][Bl]{ $W(v)$}
\begin{center}
\includegraphics[scale=0.35]{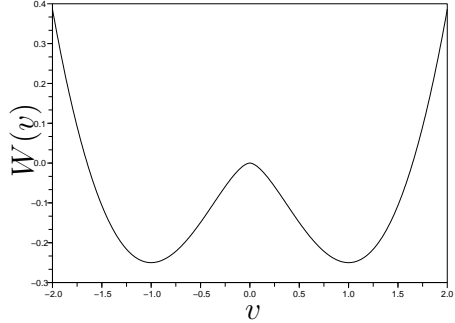}
\end{center}
\caption{\label{graphew}
Graph of the double-well logarithmic potential $W$.}
\end{figure}

The potential $W$ in (\ref{slognls})
has a double-well structure with a local maximum at $v=0$ (see figure \ref{graphew}), hence
there exists a pair of (symmetric) homoclinic orbits to $0$ and
a continuum of periodic orbits. The homoclinic solutions have the explicit form
$v(\xi )=\pm v^0 (\xi )$ with
\begin{equation}
\label{homlog}
v^0(\xi )= \sqrt{e}\, e^{-\frac{\xi^2}{4}}.
\end{equation}
Note that these (Gaussian) homoclinic solutions decay super-exponentially, but
do not decay doubly-exponentially unlike the solitary wave solutions of
the differential advance-delay equation (\ref{diffadvdel-eq})
\cite{english,stef}.

\begin{figure}[h]
\psfrag{z}[1][Bl]{ $z$}
\psfrag{w}[1][Bl]{ $w(z)$}
\begin{center}
\includegraphics[scale=0.3]{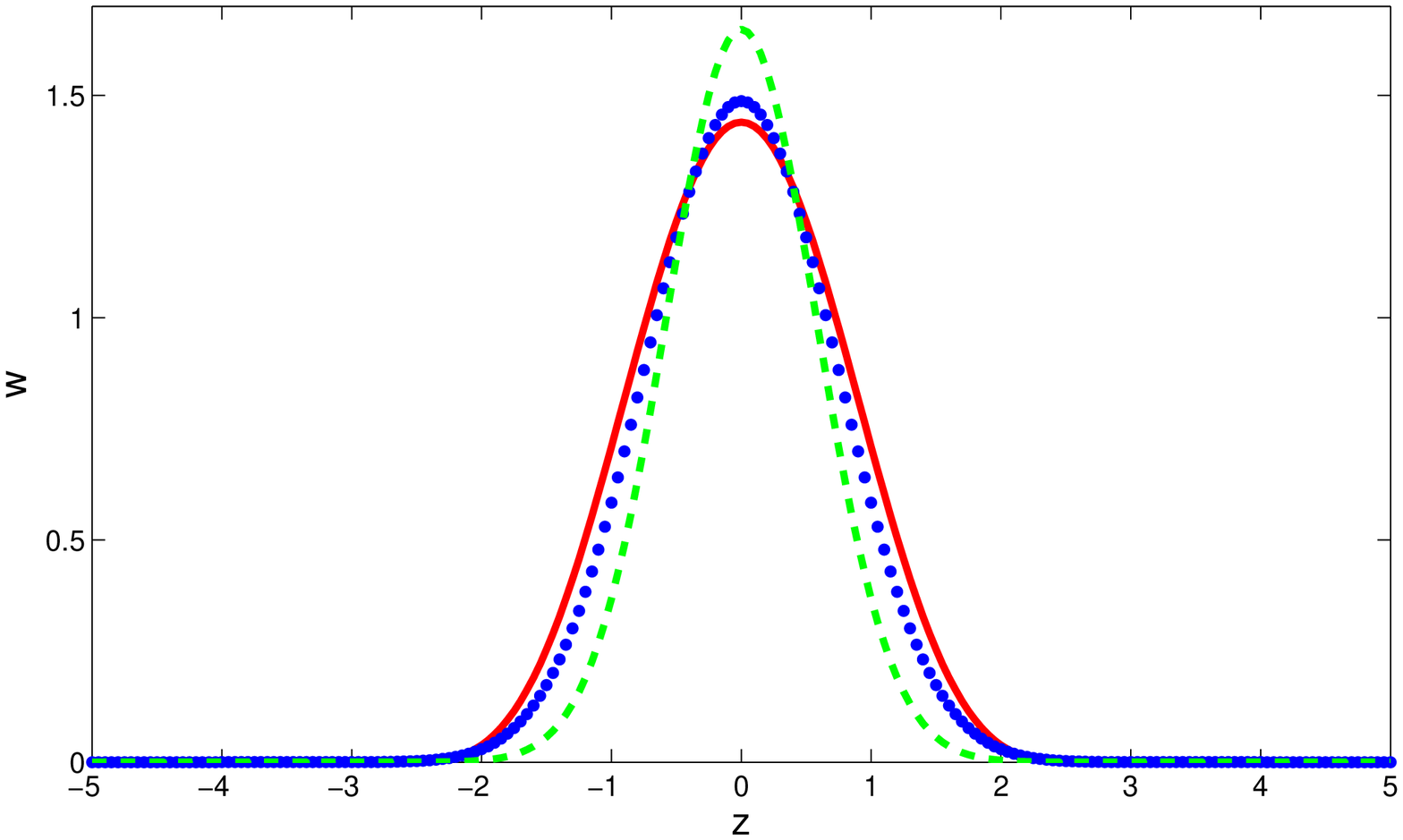}
\includegraphics[scale=0.3]{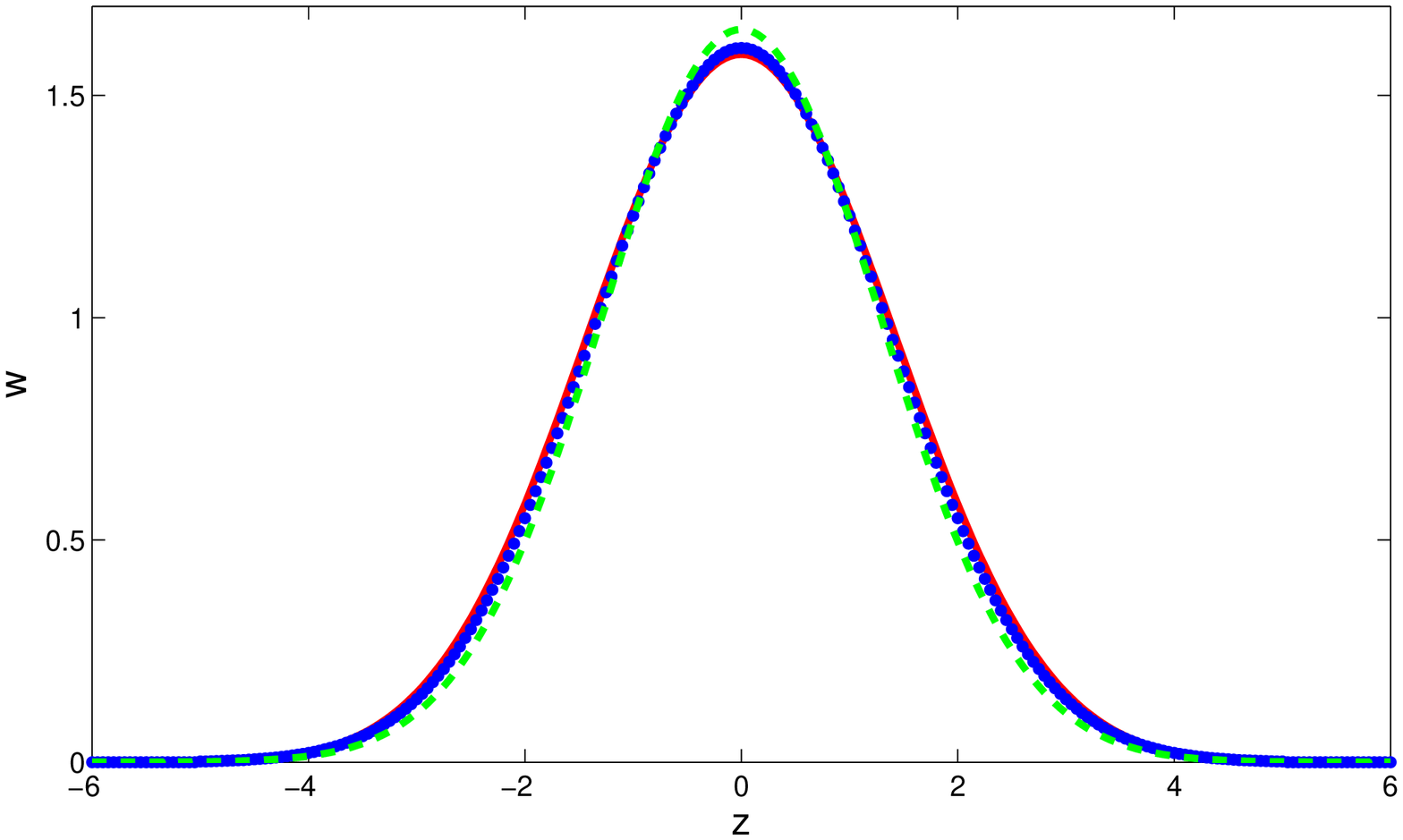}
\end{center}
\caption{\label{Soliton}
Solitary wave solution (blue dotted line) of the differential advance-delay equation (\ref{diffadvdel-eq})
in comparison with the compactly supported solitary wave (red solid line)
given by (\ref{compacton}) and the Gaussian solitary wave (green dashed line) given by
(\ref{rgauss})
for $\alpha = 1.5$ (left) and $\alpha = 1.1$ (right). }
\end{figure}

\begin{figure}[h]
\psfrag{a}[1][Bl]{ $\alpha$}
\psfrag{Errorr}[1][Bl]
{\begin{turn}{-90}
$\displaystyle{\frac{\| w- w_{\rm{approx}}\|_{\infty}}{\| w\|_{\infty}}}$\hspace*{13ex}
\end{turn}}
\begin{center}
\includegraphics[scale=0.37]{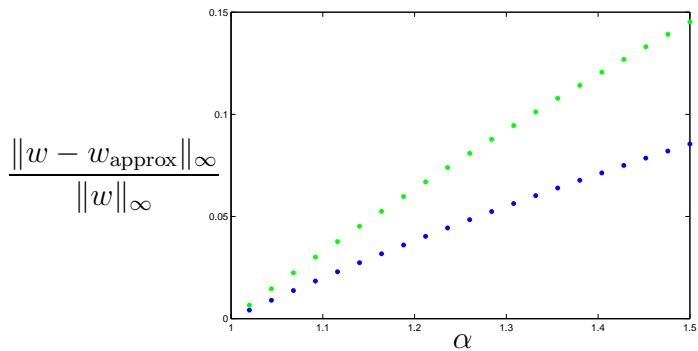}
\end{center}
\caption{\label{Convergence}
Relative error (in supremum norm)
between the solitary wave solution $w$ of the differential advance-delay equation (\ref{diffadvdel-eq})
and either the compactly supported solitary wave $w_{\rm{approx}}=w_{\rm{c}}$ given by (\ref{compacton})
(blue dots) or the Gaussian solitary wave $w_{\rm{approx}}=w_{\rm{G}}$ defined by (\ref{rgauss}) (green dots).
Both approximation errors converge to zero
as $\alpha  \rightarrow 1^+$ (absolute errors converge to zero similarly).
}
\end{figure}

The homoclinic solution (\ref{homlog})
yields an approximate Gaussian solitary wave
solution of the FPU lattice (\ref{ncrel})
with velocity equal to unity
\begin{equation}
\label{approxsolvone}
u_n(t) \approx w_{\rm{G}}(n-t),
\end{equation}
with
\begin{equation}
\label{rgauss}
w_{\rm{G}}(z):=v^0 (2\sqrt{3}\epsilon\, z)=e^{\frac{1}{2}-3 (\alpha -1)\, z^2}.
\end{equation}
Figure \ref{Soliton} compares the solitary wave solution of the differential advance-delay equation
(\ref{diffadvdel-eq}) computed numerically with
the analytical approximations corresponding to the compactly supported solitary wave
(\ref{compacton}) and the Gaussian solitary wave (\ref{rgauss}).
The numerical approximations of solitary wave solutions of (\ref{diffadvdel-eq})
were obtained using the algorithm
described in \cite{ap}, based on a reformulation of (\ref{diffadvdel-eq}) as a nonlinear
integral equation and the method of successive approximations
(see also \cite{english,herrmann} for variants of this method).
Figure \ref{Convergence} shows the relative error
(in $L^{\infty}$ norms) between solitary wave solutions of the
differential advance-delay equation (\ref{diffadvdel-eq}) and the two approximations
(\ref{compacton}) and (\ref{rgauss})
as a function of $\alpha$. The Gaussian solitary wave provides a worse approximation compared to
the compactly supported solitary wave, but both approximation errors converge to zero
as $\alpha  \rightarrow 1^+$. We have in addition 
$\lim_{\alpha\rightarrow 1^+}\| w \|_{\infty}=\| w_{\rm{G}} \|_{\infty}=\sqrt{e}$, hence
the absolute errors
between the exact solitary wave $w$ and the two approximations 
converge to zero similarly to the relative errors plotted in figure \ref{Convergence}.

So far, we have computed a solitary wave solution of
the FPU lattice (\ref{ncrel}) with unit velocity and have
checked its convergence towards the Gaussian approximation (\ref{approxsolvone})
when $\alpha \rightarrow 1^+$. We shall now examine the convergence of
solitary waves with velocities different from unity.
Using the Galilean invariance of (\ref{logkdv}), the homoclinic solution
(\ref{homlog}) yields two (symmetric)
families of solitary wave solutions of
the log-KdV equation (\ref{logkdv})
\begin{equation}
\label{sol}
v(\xi ,\tau )=\pm e^{\frac{1}{2}+\tilde{c}}\, e^{-\frac{(\xi - \tilde{c}\, \tau)^2}{4}}
\end{equation}
parameterized by the wave velocity $\tilde{c}$. These profiles
yield the approximate solitary wave
solutions of the original FPU lattice (\ref{ncrel})
\begin{equation}
\label{approxsol}
u_n(t)\approx \pm\, e^{2c + \frac{1}{2}  - 3\,(\alpha -1) (n-v_{\rm{s}}\, t-\xi_0\,)^2},
\end{equation}
where we have set $\tilde{c}=2\, c$,
\begin{equation}
\label{velform}
v_{\rm{s}}=1+c\, (\alpha -1),
\end{equation}
introduced an additional phase shift $\xi_0 \in \mathbb{R}$
and used the fact that $\epsilon = \sqrt{ \alpha -1 }$.
One can observe that
the width of the approximate solitary wave
(\ref{approxsol}) diverges as $(\alpha -1)^{-1/2}$ when $\alpha \rightarrow 1^+$.
Moreover, similarly to solitary wave solutions of the FPU lattice (\ref{ncrel}),
the wave width remains constant if $\alpha$ is fixed and the wave amplitude
$a=\exp{(\frac{1}{2}+2\, c)}$ (or equivalently the wave velocity $v_{\rm{s}}$)
is varied.
In addition,
approximation (\ref{approxsol}) can be rewritten
\begin{equation}
\label{approxsolbis}
u_n(t)\approx \pm\, e^{2c}\, w_{\rm{G}}(n-v_{\rm{s}}\, t-\xi_0\,),
\end{equation}
where the renormalized Gaussian profile $w_{\rm{G}}$ is
defined in (\ref{rgauss}).
One can notice that ${|v_{\rm{s}}|}^{\frac{2}{\alpha -1}}=e^{2c}+\mathcal{O}( |\alpha -1 |)$
if $c \in \mathbb{R}$ is fixed $\alpha \rightarrow 1^+$.
Consequently, approximation (\ref{approxsolbis}) is close to a rescaling of
(\ref{approxsolvone}) through the invariance
$u_n \mapsto {| v_{\rm{s}} |}^{\frac{2}{\alpha -1}}\, u_n(v_{\rm{s}} t)$
of (\ref{ncrel}). This observation illustrates why the (nonstandard)
Galilean invariance (\ref{scaling}) of the log-KdV equation is inherited
from the scaling invariance of the FPU lattice (\ref{ncrel}).

Similarly, the solitary wave solution $w$ of the differential
advance-delay equation (\ref{diffadvdel-eq})
yields the family of solitary wave solutions of (\ref{ncrel})
\begin{equation}
\label{exactvqqe}
u_n (t) = \pm \, {|v_{\rm{s}}|}^{\frac{2}{\alpha -1}}\, w(n-v_{\rm{s}}\, t - \xi_0) .
\end{equation}
One can distinguish two contributions to the error between the Gaussian
approximation (\ref{approxsolbis}) and the exact solution (\ref{exactvqqe}),
one originating from the profile function $w_{\rm{G}}$ and the other
from the wave amplitude.
From the numerical results of figure \ref{Convergence}, we know
that $\| w - w_{\rm{G}}\|_\infty \rightarrow 0$ when $\alpha \rightarrow 1^+$.
In addition, fixing $c \in \mathbb{R}$ and considering wave velocities (\ref{velform})
close to unity when $\alpha \rightarrow 1^+$, we have seen that
$\lim_{\alpha \rightarrow 1^+}{|v_{\rm{s}}|}^{\frac{2}{\alpha -1}}=e^{2c}$.
Consequently, the exact solitary wave (\ref{exactvqqe}) with velocity (\ref{velform})
converges uniformly towards the Gaussian approximation (\ref{approxsolbis})
when $\alpha \rightarrow 1^+$.

\begin{remark}
Note that the
convergence result above concerns solitary waves
with velocities converging towards unity, i.e. the value of sound velocity
in the linear chain with $\alpha =1$.
This restriction is due to the specific scaling (\ref{scal}) assumed for solutions
described by the log-KdV equation.
On the contrary, the exact FPU solitary wave (\ref{exactvqqe}) with fixed velocity
$v_{\rm{s}} \neq \pm 1$ becomes degenerate when $\alpha \rightarrow 1^+$,
since the wave amplitude
goes to $0$ if $|v_{\rm{s}}|<1$ and diverges if $|v_{\rm{s}}|>1$.
\end{remark}

\begin{remark}
The compactly supported solitary wave (\ref{compacton}) yields approximate solutions
$u^{\rm{c}}$ of the FPU system (\ref{ncrel}) taking the form
$u_n^{\rm{c}} (t) = \pm \, {|v_{\rm{s}}|}^{\frac{2}{\alpha -1}}\, w_{\rm{c}}(n-v_{\rm{s}}\, t - \xi_0)$.
This approximation can be compared to the exact FPU solitary wave $u$ defined by
(\ref{exactvqqe}). The results of figure \ref{Convergence} show that
$\| w -  w_{\rm{c}} \|_{\infty} \to 0 $ as $\alpha \to 1^+$. Consequently,
one can infer that the relative error 
$\| u^{\rm{c}} - u  \|_{L^\infty (\mathbb{Z}\times \mathbb{R})}
/\| u   \|_{L^\infty (\mathbb{Z}\times \mathbb{R})}$
converges to zero
when $v_{\rm{s}}$ is fixed and $\alpha \rightarrow 1^+$.
\end{remark}

\subsection{\label{gauss}Formation of Gaussian solitary waves}

In section \ref{trav}, we have computed solitary wave solutions of
the FPU lattice (\ref{ncrel}) and have
checked their convergence towards the Gaussian approximation (\ref{approxsol})
when $\alpha \rightarrow 1^+$. These results are valid in a stationary regime
and for prescribed wave velocities converging towards unity. To complete this analysis, we shall
study the formation of solitary waves from a localized perturbation of given
magnitude, and compare their profiles to Gaussian approximations when $\alpha$
is close to one.

In what follows, we numerically integrate the FPU
system (\ref{nc}) with Hertzian potential (\ref{vhertz})
for different values of $\alpha >1$.  We consider the Hertzian potential (\ref{vhertz}) rather that the symmetrized potential
(\ref{vhom}) because of its relevance to impact mechanics. In addition,
the differential equations (\ref{nc}) are easier to integrate numerically
due to the absence of dispersive wavetrains for the above initial condition.

We consider a lattice of $N=2000$ particle with free-end boundary conditions.
Computations are performed with
the standard ODE solver of the software Scilab.
We consider a velocity perturbation of the first particle (at $n=0$), corresponding to
the initial condition
\begin{equation}
\label{ic}
x_n(0)=0 \
\text{ for~all } \
n \geq 0, \ \ \
\dot{x}_0 (0)=0.1, \ \ \
\dot{x}_n (0)=0 \
\text{ for~all } \
n \geq 1.
\end{equation}
Due to the scale invariance of system (\ref{nc}) and (\ref{vhertz}), all positive initial velocities
for $\dot{x}_0(0)$ yield a rescaled solution of the form
$\tilde{x}_n (t)= v^{\frac{2}{\alpha +1}}\, x_n ( v^{\frac{\alpha -1}{\alpha +1}}\, t)$.

The front edge of the solution evolves into a solitary wave
whose profile becomes stationary for large enough times, at least on the timescales of the simulations.
When $\alpha$ is sufficiently close to unity, one observes that the asymptotic velocity of
the solitary wave is close to unity (this property is also true for any initial velocity $\dot{x}_0 (0)>0$).
The solitary wave is of compression type (i.e. with $u_n <0$), hence it also defines a solution of
the FPU lattice (\ref{ncrel}) with the symmetrized potential (\ref{vhom})
and can be compared with approximation (\ref{approxsol}).

\begin{figure}[h]
\psfrag{n}[1][Bl]{ $n$}
\psfrag{u_n}[1][Bl]{ $u_n (t)$}
\begin{center}
\includegraphics[scale=0.35]{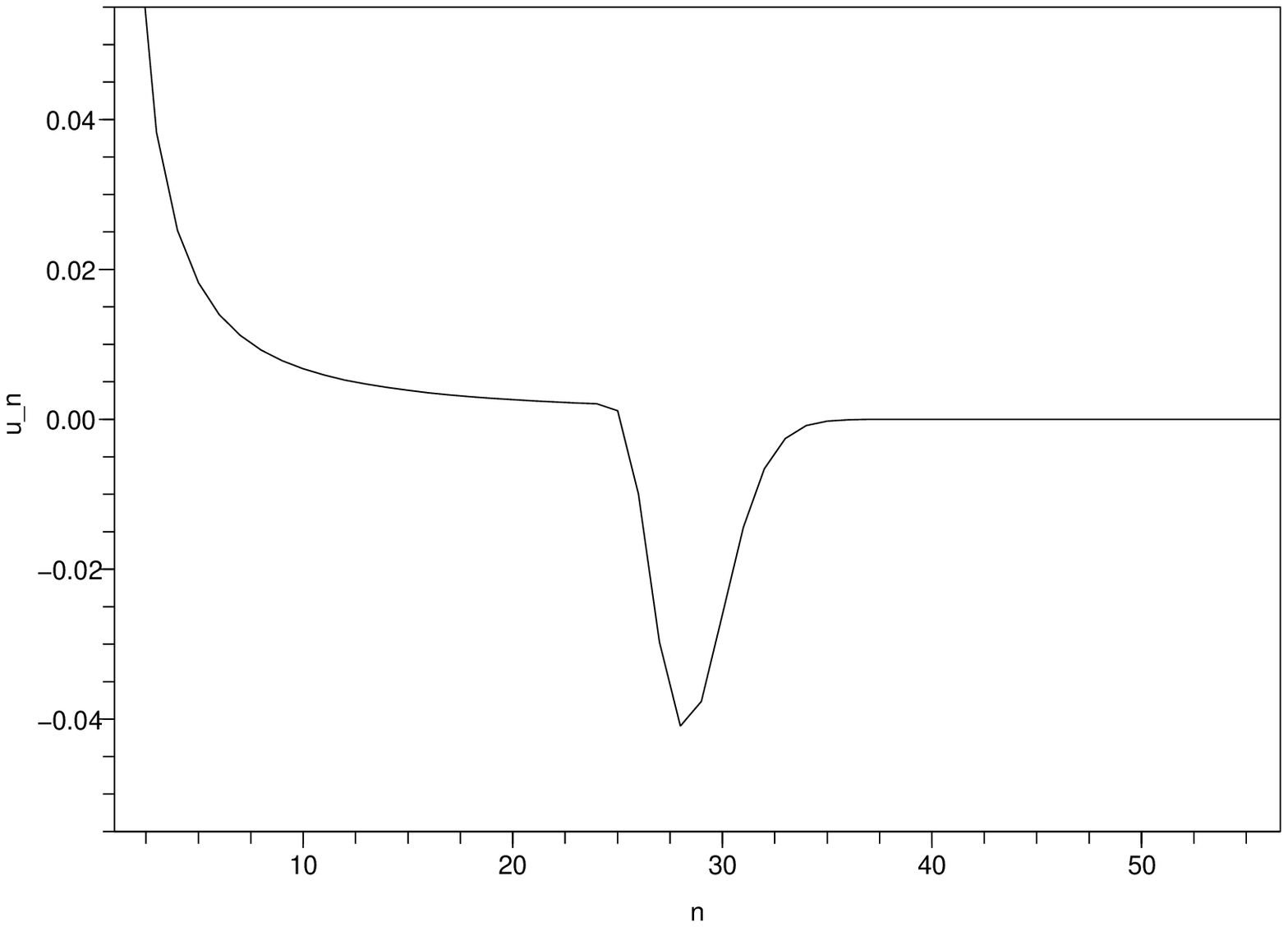}
\includegraphics[scale=0.35]{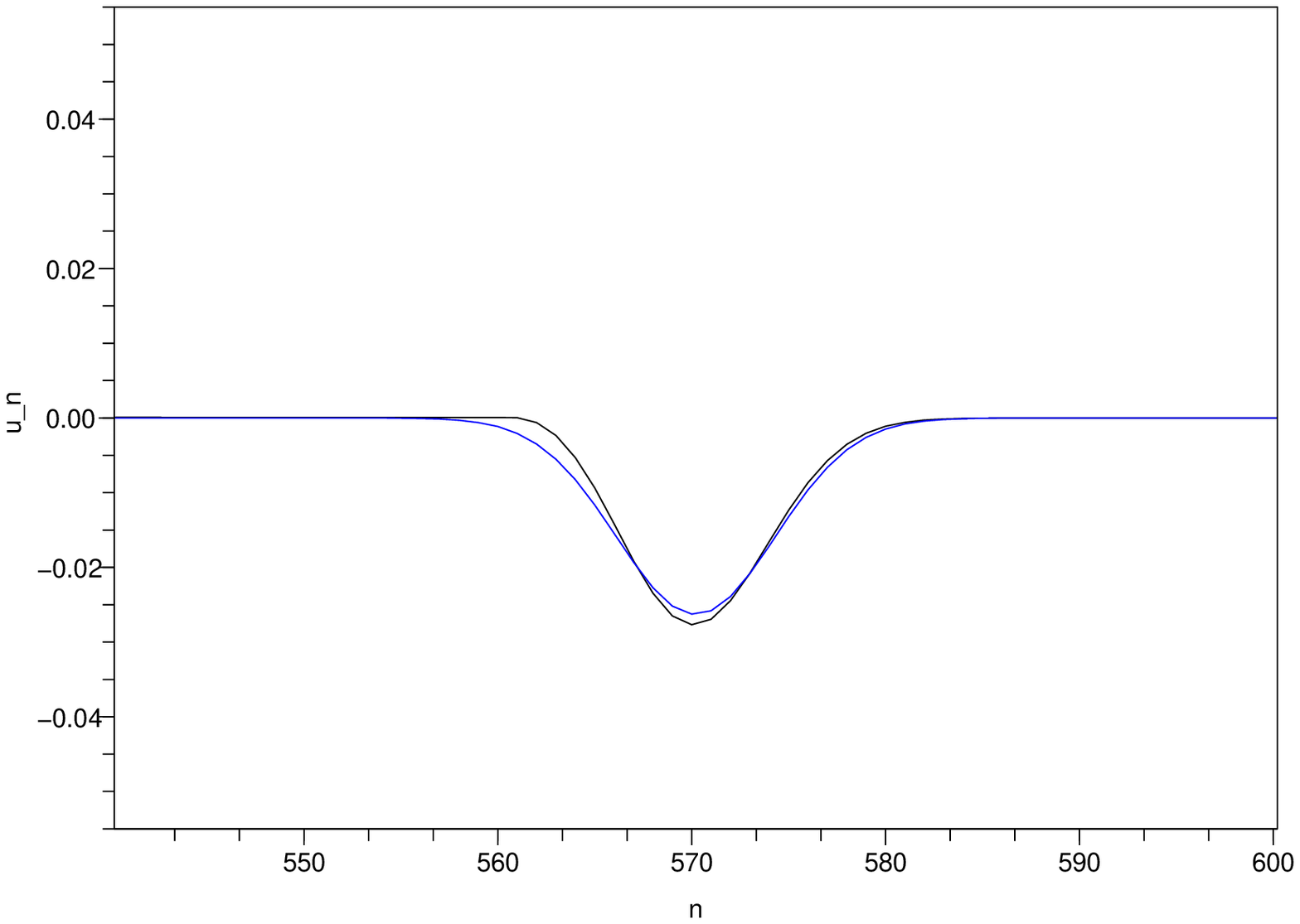}
\includegraphics[scale=0.35]{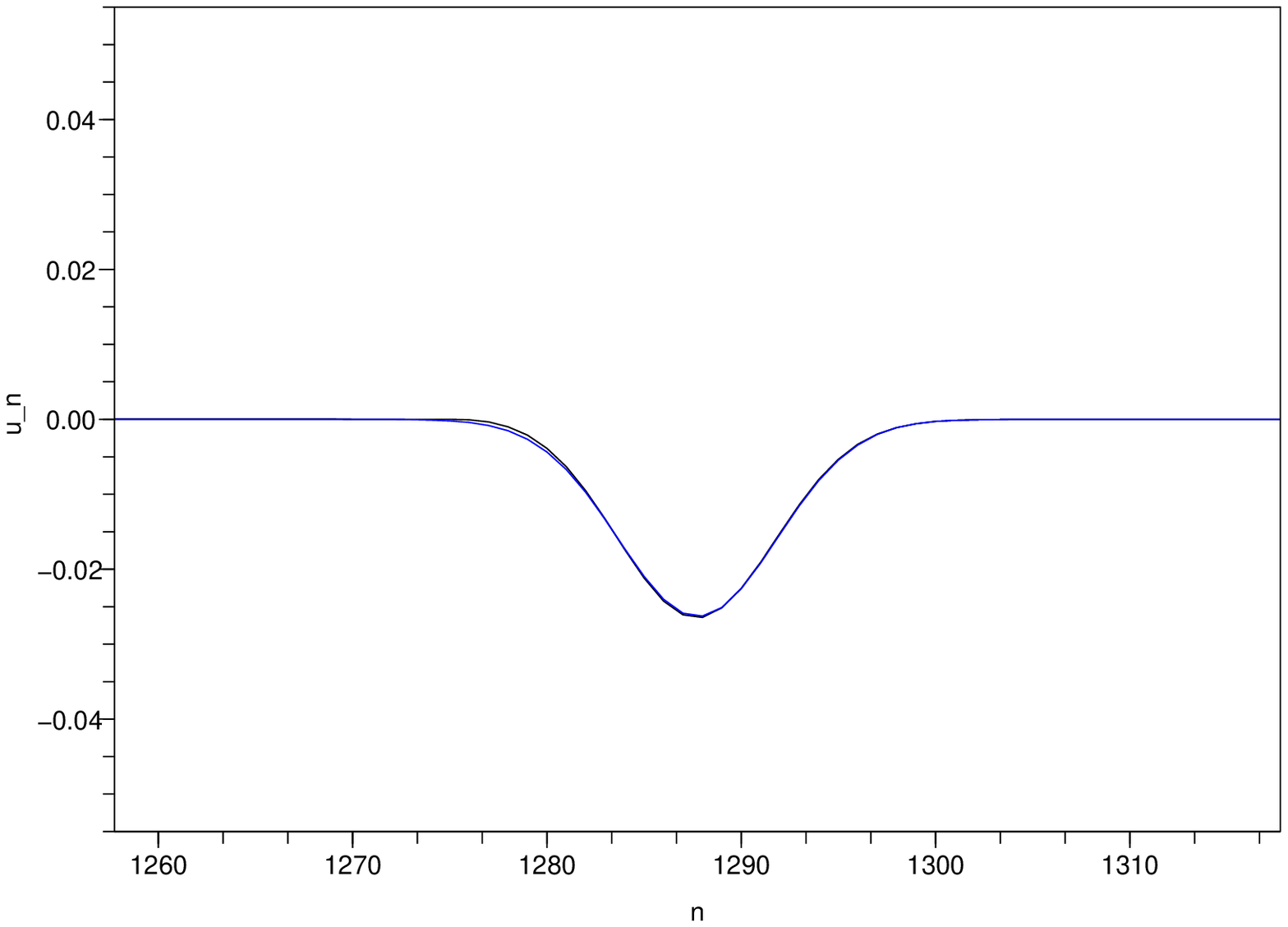}
\end{center}
\caption{\label{pulse1_01}
Formation of a localized wave resulting from the initial condition (\ref{ic})
in the Hertzian FPU lattice (\ref{nc}) and (\ref{vhertz}) with $\alpha = 1.01$.
Top left~: solution profile $u_n$ at $t\approx 30.5$ (note the absence of dispersive
wavetrains behind the localized pulse). Top right~:
the pulse profile at $t\approx 585.6$ (black curve)
is compared with the Gaussian approximation (\ref{approxsol}) (blue curve).
A small mismatch between the two curves is visible at this stage
(note the slightly asymmetric shape of the numerical solution).
The value of $c$ has been fixed so that
the exact and approximate solitary waves
have the same amplitude for large values of $t$.
Bottom~: same comparison at $t\approx 1318.3$, where
the approximate and numerical solutions almost perfectly coincide.
}
\end{figure}

The results are shown in figure \ref{pulse1_01} for $\alpha = 1.01$.
The parameter $c\approx -2.07$ in the Gaussian approximation (\ref{approxsol})
is determined from the relation $a=\exp{(\frac{1}{2}+2\, c)}$, where
$a$ is the exact solitary wave amplitude obtained by integrating
(\ref{nc}) with initial data (\ref{ic}).
The approximation of the solitary wave profile is very accurate
in the stationary regime, as shown by the bottom panel. In addition, the measured velocity
of the numerical solitary wave solution $v_{\rm{num}} \approx 0.9789$ can be
compared with the velocity $v_{\rm{app}}=1+c\, (\alpha -1) \approx 0.9793$ of the approximate
solitary wave (\ref{approxsol}), which yields
a relative error $E=|v_{\rm{app}}-v_{\rm{num}}|/v_{\rm{num}}$ around $0.04\%$.

Discrepancies appear between the profiles of the
numerical solution and the Gaussian approximation for larger values of $\alpha$,
as already noticed in figures \ref{Soliton} and \ref{Convergence}. In addition, we find
a relative error $E$ between numerical and approximate wave velocities
around $7\%$ for $\alpha = 1.222$ and $36\%$ for $\alpha = 1.5$.
As a conclusion, while
some quantitative agreement is still obtained for $\alpha \approx 1.2$
between the numerical solution and the Gaussian approximation, the latter becomes
unsatisfactory for $\alpha = 1.5$.

\subsection{\label{lstab}Linear stability of Gaussian solitary waves}

The numerical results of section \ref{gauss} indicate the
long time stability of the solitary wave solutions of
system (\ref{nc}) with Hertzian potential (\ref{vhertz})
that form after a localized perturbation. Therefore,
the stability of the Gaussian solitary waves (\ref{sol})
appears as a necessary (of course not sufficient)
condition to establish the validity of the log-KdV equation (\ref{logkdv})
as a modulation equation for the FPU system with Hertzian potential.
In this section, we prove
the linear orbital stability
of solitary waves of the log-KdV equation.
We perform the analysis for the stationary
Gaussian solution $v^0$ defined by (\ref{homlog}).
By the scaling transformation (\ref{scaling}), the stability result
extends to the entire family of solitary waves
$e^c\, v^0(\xi - c\, \tau )$ with $c\in \mathbb{R}$.

The log-KdV equation (\ref{logkdvpot}) can be written in the Hamiltonian form
\begin{equation}
\label{kdvham}
\partial_{\tau} v = \partial_{\xi} E'(v),
\end{equation}
associated with the energy (\ref{energy}).
The stationary Gaussian solution $v^0$
is a critical point of the energy $E(v)$, i.e. $E'(v^0) = 0$.
The Hessian operator evaluated at this solution reads
\begin{equation}
L := E''(v^0) = -\partial_{\xi}^2 - 1 - \log|v^0| = -\partial_{\xi}^2 - \frac{3}{2} + \frac{\xi^2}{4}
\end{equation}
and corresponds to
a Schr\"{o}dinger operator with a harmonic potential.
Equation (\ref{kdvham}) linearized at $v^0$ reads
\begin{equation}
\label{lkdvham}
\partial_{\tau} v = \partial_{\xi} L\, v.
\end{equation}
In what follows, we formulate (\ref{lkdvham})
as a differential equation in suitable function spaces and
derive a linear stability result based on the energy method
for KdV-type evolution equations \cite{bona}.

The spectral properties of $L$ are well known \cite{shubin}.
The operator $L$ is self-adjoint in $L^2(\mathbb{R})$ with dense domain
$$
D(L) = \{\, u\in H^2(\mathbb{R}), \, \xi^2\, u \in L^2(\mathbb{R})\, \}.
$$
Its spectrum consists of simple eigenvalues at
integers $n - 1$, where $n \in \mathbb{N}_0$ (the set of natural numbers including zero).
In particular, $L$ has a simple eigenvalue $-1$ with eigenspace spanned by $\phi_0 = v^0 / \| v^0 \|_2$,
a simple zero eigenvalue with eigenspace spanned by $\phi_0^\prime =\partial_\xi \phi_0$, and
the rest of its spectrum is bounded away from zero by a positive number.
The discreteness of the spectrum comes from the fact that
the harmonic potential of the Schr\"{o}dinger operator $L$ is unbounded
at infinity, which implies that
the embedding $D(L)\hookrightarrow L^2(\mathbb{R})$ is compact
(see e.g. \cite{shubin}, p. 43-44) and $L$ has a compact resolvent in $L^2(\mathbb{R})$.

Hereafter we denote by
$\left( \cdot,\cdot \right) $ the usual $L^2$-scalar product.
The operator $\partial_{\xi} L\, : \, D(L) \rightarrow H^{-1}(\mathbb{R})$
inherits a double non semi-simple zero eigenvalue,
with generalized kernel $E_0$ spanned by
the eigenvector $\phi_0^\prime$ and
the generalized eigenvector $\phi_0$.
The algebraic multiplicity of this eigenvalue is $2$ because equation
$\partial_{\xi} L\, u = \phi_0$ has no solution in $D(L)$
(since $\langle \partial_{\xi} L\, u , \phi_0  \rangle = -( u , L\,  \phi_0^\prime  )=0 \neq \| \phi_0 \|_2^2$).
The double zero eigenvalue is linked with
the existence of a two-parameter family of solitary waves of the log-KdV equation (\ref{logkdv}) parameterized by
the location and velocity of the waves. It induces in general
a secular growth of the solutions of (\ref{lkdvham}) along the eigenvector $\phi_0^\prime$, linked
with a velocity change of perturbed solitary waves. In order to prove a linear stability result,
we thus have to project (\ref{lkdvham}) onto the invariant subspace under $\partial_{\xi} L$
associated with
the nonzero part of the spectrum. Following a classical computation scheme \cite{pw},
the spectral projection onto $E_0$ takes the form $P_0\, v = a\, \phi_0^\prime + b\, \phi_0$ with
$$
a = -\left( v , \int_0^\xi{\phi_0\, dx}  \right) , \ \ \
b =  \left( v , \phi_0  \right)
$$
(one can readily check that $P_0$ commutes with $\partial_{\xi} L$).
Note in passing that $a$ is well defined because $D(L) \subset L^1 (\mathbb{R})$.
Now, splitting the solutions $v(\cdot ,\tau ) \in D(L)$ of (\ref{lkdvham}) into
$$
v(\xi ,\tau )=a(\tau )\, \phi_0^\prime(\xi) + b(\tau )\, \phi_0 (\xi ) + y(\xi , \tau )
$$
with $y(\cdot,\tau ) = (I-P_0)\, v(\cdot,\tau )$,
one obtains the following equivalent system
\begin{equation}
\label{eq0}
\frac{d a}{d \tau }=-b, \ \ \ \frac{d b}{d \tau }=0,
\end{equation}
\begin{equation}
\label{eq1}
\partial_{\tau} y = \partial_{\xi} L\, y.
\end{equation}
In order to prove the linear (orbital) stability of the Gaussian solitary wave, we have to show
the Lyapunov stability of the equilibrium $y=0$ of the linear evolution equation
(\ref{eq1}) for a suitable topology. 
Let us recall that $P_0 y(\cdot,\tau )=0$, hence
$y(\cdot,\tau )$ belongs to the codimension-$2$ subspace of $D(L)$ defined as
$$
D_1 = \left\{ \,  y\in D(L): \quad   \left( y , \phi_0  \right) = \left( y , \int_0^\xi{\phi_0\, dx}  \right) =0    \, \right\} .
$$
Since $L$ is positive on ${\phi_0}^\perp$ and $D_1 \subset {\phi_0}^\perp$, we  can define
$\| y \|_{L} = (L\, y , y)^{1/2}$. Due to the fact that
$L$ is positive-definite on ${\phi_0}^\perp \cap {\phi_0^\prime}^\perp$ and
$\phi_0^\prime \notin D_1$, $\| \cdot \|_{L} $ defines a norm on $D_1$
(roughly a weighted $H^1$-norm).
Denote by $H_1$ the completion of $D_1$ with respect to the norm $\| \cdot \|_{L}$.
This norm defines a convex conserved
Lyapunov function for system (\ref{eq1}) since
\begin{equation}
\label{conserv}
\frac{d}{d\tau}\, \frac{1}{2}\, \| y \|_{L}^2 = (L\, y , \partial_\tau y) = (L\, y , \partial_\xi L y) = 0.
\end{equation}
For simplicity, let us choose an initial data in the Schwartz space
$\mathcal{S}(\mathbb{R})$ of rapidly decreasing $C^\infty$ functions:
\begin{equation}
\label{cilin}
y_{|\tau =0}=y_0 \in D_1 \cap \mathcal{S}(\mathbb{R}).
\end{equation}
Thanks to property (\ref{conserv}), we get from the energy method \cite{rg}
and standard bootstrapping arguments (see e.g. chapter 11.1.4 of \cite{rg})
a unique global solution of (\ref{eq1})-(\ref{cilin})
which is infinitely smooth in time and space.
We have $y \in L^\infty (\mathbb{R},H_1)$ and
$\| y(\cdot , \tau ) \|_{L} = \| y_0 \|_{L}$
for all $\tau \in \mathbb{R}$, which shows
the Lyapunov stability of the equilibrium $y=0$ of (\ref{eq1}) in $H_1$.
Therefore, we have
proved the linear orbital stability of the Gaussian solitary wave.

The Lyapunov stability of the equilibrium $y=0$
implies the absence of eigenvalue of $\partial_\xi L$ with positive real part, since
such an eigenvalue would lead to exponential growth of the solution along a corresponding eigenvector.
In addition, the spectrum of $\partial_{\xi} L$ is invariant by $\lambda \rightarrow -\lambda$
since $\partial_{\xi} L$ possesses a reversibility symmetry, i.e.
anticommutes with the symmetry $v(\xi) \mapsto v(-\xi )$. This implies that
all the eigenvalues of $\partial_{\xi} L$ lie on the imaginary axis.
This result contrasts with the instability of
the compacton solutions of the fully nonlinear Boussinesq equation
numerically analyzed in section \ref{bouss}. Moreover, it is consistent with
the absence of solitary wave instabilities observed in
section \ref{gauss}.

\begin{remark}
The absence of eigenvalues of $\partial_\xi L$ with positive real part could be also obtained from
the recent works \cite{kapstef,dmitrystab}. This result follows from
the main theorems in \cite{kapstef,dmitrystab} if
the number of negative eigenvalues of $L$ is equal to one and
\begin{equation}
\label{last-assumption}
( L^{-1} \partial_{\xi}^{-1} \phi_0^\prime, \partial_{\xi}^{-1} \phi_0^\prime )  < 0 ,
\end{equation}
where $\partial_{\xi}^{-1} u := \int_{-\infty}^\xi{u\, dx}$.
Assumption (\ref{last-assumption}) is satisfied since
$\partial_{\xi}^{-1} \phi_0^\prime =\phi_0 \in \text{Range}(L)$
(recall $L\, \phi_0=-\phi_0$) and
$
( L^{-1} \partial_{\xi}^{-1} \phi_0^\prime, \partial_{\xi}^{-1} \phi_0^\prime ) =
( L^{-1} \phi_0, \phi_0 ) =  - \| \phi_0 \|_2^2 < 0 .
$
\end{remark}

\section{\label{comprev}Compacton approximation revisited}

The results of figures \ref{Soliton} and \ref{Convergence} indicate that compacton approximations
converge towards solitary wave solutions of the differential advance-delay equation
(\ref{ncrel}) when $\alpha \rightarrow 1^+$.
It seems delicate to establish this result directly from the methodology described in section \ref{bouss},
where equation (\ref{compacton-eq}) is heuristically obtained by
truncation of the differential advance-delay equation (\ref{diffadvdel-eq}).
However, it is instructive to compare analytically the compacton approximation
$w_{\rm{c}}(z)$ defined in (\ref{compacton})
and the Gaussian approximation (\ref{rgauss}) when $\alpha$ is close to unity,
since our numerical results indicate that both profiles become very close
(see figure \ref{Soliton}, right panel).
To check the consistency of (\ref{compacton}) and (\ref{rgauss}),
we note that the compact support $\left[-\frac{\pi}{2 B},\frac{\pi}{2 B}\right]$ of approximation (\ref{compacton})
extends to the entire real line as $\alpha \rightarrow 1^+$. Furthermore,
let $\epsilon := \sqrt{\alpha - 1}$, $\xi := 2 \sqrt{3} \epsilon z$,
and perform the expansions
\begin{equation}
\label{expa}
A = \left( \frac{2+\epsilon^2}{2 (1 + \epsilon^2)} \right)^{-\frac{1}{\epsilon^2}} =
e^{-\frac{1}{\epsilon^2} \log\left( 1- \frac{\epsilon^2}{2(1+\epsilon^2)}\right)} =
\sqrt{e} \left(1 + \mathcal{O}(\epsilon^2) \right)
\end{equation}
and
\begin{equation}
\label{expb}
\cos^{\frac{2}{\alpha -1}}(B z) = e^{\frac{2}{\epsilon^2} \log \cos\left(\frac{\epsilon \xi}{2( 1 + \epsilon^2)} \right)} =
e^{\frac{2}{\epsilon^2} \log \left(1 - \frac{\epsilon^2 \xi^2}{8(1 + \epsilon^2)^2} + \mathcal{O}(\epsilon^4 \xi^4) \right)}
= e^{-\frac{\xi^2}{4}} \left( 1 + \mathcal{O}(\epsilon^2 ) \right)
\end{equation}
for all fixed $\xi \in \mathbb{R}$. From expansions (\ref{expa}) and (\ref{expb}),
it follows that the renormalized compacton
$w_{\rm{c}}(\xi / \sqrt{12 (\alpha -1 )})$ converges towards the Gaussian solution (\ref{homlog})
for any fixed $\xi \in \mathbb{R}$ when $\alpha \rightarrow 1^+$.

One possible approach for the justification of compactons
consists in deriving
an asymptotic model consistent with FPU lattice (\ref{ncrel})
and supporting compacton solutions, in analogy with the
derivation of the log-KdV equation. Such a model may be free of
the artificial instabilities introduced by the Boussinesq equation (\ref{B}),
and may lead to a well-posed evolution problem in a suitable
function space and for appropriate initial data.

In what follows we show that a generalized KdV equation with
nonsmooth fractional power nonlinearity can be derived from (\ref{ncrel})
using the method of section \ref{kdvloga}.
For this purpose, we note that the nonlinearity
$f_\alpha (u) $ defined by (\ref{deffa})
satisfies
\begin{equation}
\label{deffam1}
f_\alpha (u)
= (\alpha -1)\, u \ln{|u|} + \mathcal{O}((\alpha -1)^2)
= \alpha \left( u-u\, |u|^{\frac{1}{\alpha} -1} \right) + \mathcal{O}((\alpha -1)^2) 
\end{equation}
(uniformly in $u$ on bounded intervals)
when $\alpha \rightarrow 1^+$.
Consequently, the generalized KdV equation
\begin{equation}
\label{rac}
\partial_\tau v + \partial^3_\xi v + \frac{\alpha}{\alpha -1}\, \partial_\xi ( v-v\, |v|^{\frac{1}{\alpha} -1} )=0
\end{equation}
is consistent with system (\ref{ncrel})
at the same order as the log-KdV equation,
and converges towards the log-KdV equation when $\alpha \rightarrow 1^+$.
We call equation (\ref{rac}) the H\"olderian KdV (H-KdV) equation due to the presence
of the H\"older-continuous nonlinearity $v\, |v|^{\frac{1}{\alpha} -1}$.
This approximation of the logarithmic nonlinearity is reminiscent of
results of \cite{js12} obtained for a stationary H\"olderian NLS equation close to a logarithmic limit.

Note that the H-KdV equation (\ref{rac}) admit the same three conserved quantities
(\ref{mass}), (\ref{momentum}), and (\ref{energy}) with potential $W$ replaced by
$$
\tilde{W}(v) = \frac{\alpha}{\alpha - 1} \left( \frac{1}{2} v^2 - \frac{\alpha}{\alpha+1} |v|^{1+\frac{1}{\alpha} }\right).
$$
Moreover, equation (\ref{rac}) admits a nonstandard Galilean invariance similarly
to the log-KdV equation. More precisely,
any solution $v$ of (\ref{rac}) generates a one-parameter family of solutions
$\phi(c) \cdot v$ defined by
$$
\left[ \phi(c) \cdot v\right](\xi,\tau) := (1-\mu )^{\frac{\alpha}{1-\alpha}}
v\left((1- \mu )^{\frac{1}{2}}(\xi-c\, \tau),(1-\mu )^{\frac{3}{2}} \tau\right) , \ \ \
\mu= c\, \left( 1-\frac{1}{\alpha} \right),
$$
and parameterized by $c \in (-\infty , \frac{\alpha}{\alpha -1} ]$. One can check that
this symmetry reduces to the Galilean invariance (\ref{scaling}) when $\alpha \rightarrow 1^+$.

Let us check that the H-KdV equation admits compacton solutions. Their
existence is due to the non-differentiable H\"older-continuous nonlinearity,
in contrast to classical compacton equations where degenerate nonlinear dispersion
plays a central role \cite{rh}.
The stationary H-KdV equation integrated once reads
\begin{equation}
\label{racs}
\frac{d^2 v}{d\xi^2} + \frac{\alpha}{\alpha -1}\,  ( v-v\, |v|^{\frac{1}{\alpha} -1} )=0 ,
\end{equation}
(the integration constant has been set to $0$), or equivalently
$\frac{d^2 v}{d\xi^2} + \tilde{W}^\prime (v) =0$. This equation is integrable and
the potential $\tilde{W}$ has a double-well structure with a local maximum at $v=0$.
This property implies the existence of a pair of symmetric homoclinic orbits to $0$
corresponding to compactons. Using the change of variable
$$
w(z)=(v\, |v|^{\frac{1}{\alpha} -1})(\xi ), \quad z = \frac{\sqrt{\alpha} \xi}{\sqrt{12(\alpha-1)}},
$$
equation (\ref{racs}) is mapped to the form
(\ref{compacton-eq}) which possesses compacton solutions given by (\ref{compacton}).
Consequently, equation (\ref{racs}) admits stationary compacton solutions
\begin{equation}
\label{compactonkdv}
v_\alpha (\xi ) = \left\{ \begin{array}{ll} \tilde{A} \cos^{\frac{2 \alpha }{\alpha-1}}(\tilde{B} \xi ), & |\xi | \leq \frac{\pi}{2 \tilde{B}}, \\
0, & |\xi | \geq \frac{\pi}{2 \tilde{B}}, \end{array} \right.
\end{equation}
where
$$
\tilde{A} = \left( \frac{1+\alpha}{2 \alpha} \right)^{\frac{\alpha}{1-\alpha}}, \quad \tilde{B} = \frac{\sqrt{\alpha-1}}{2\sqrt{\alpha}}.
$$
In addition, we have as above $\lim_{\alpha \rightarrow 1^+}{v_\alpha (\xi )}= \sqrt{e}\,  e^{-\frac{\xi^2}{4}}$
for any fixed $\xi \in \mathbb{R}$, i.e. the compacton approximation (\ref{compactonkdv})
converges towards the Gaussian solitary wave approximation (\ref{homlog})
when $\alpha \rightarrow 1^+$.

\begin{remark}
The compacton approximation $w_{\rm{c}}(\xi / \sqrt{12 (\alpha -1 )})$ defined in (\ref{compacton}) and
obtained using the Boussinesq equation (\ref{B}) differs from the compacton (\ref{compactonkdv})
deduced from the H-KdV equation. However, both approximations are equivalent
when $\alpha \rightarrow 1^+$ since they
converge towards the Gaussian profile (\ref{homlog}).
In fact, an infinity of compacton approximations could be constructed, depending on
the approximation of the logarithmic nonlinearity introduced in (\ref{deffam1}).
\end{remark}

Using the Galilean invariance of (\ref{rac}), the compacton
(\ref{compactonkdv}) yields two (symmetric)
families of compactly supported solitary waves of
the H-KdV equation (\ref{rac})
\begin{equation}
\label{soltrav}
v(\xi ,\tau )=\pm (1-\mu )^{\alpha/(1-\alpha)}\, v_\alpha [\, \sqrt{1-\mu}\, (\xi - \tilde{c}\, \tau)  \, ],
\ \ \ \mu=\tilde{c}\, (1-\frac{1}{\alpha}),
\end{equation}
parameterized by the wave velocity $\tilde{c}  \in (-\infty , \frac{\alpha}{\alpha -1} ]$.
From the expression (\ref{scal}), these profiles
yield the approximate compacton solutions of the FPU lattice (\ref{ncrel})
\begin{equation}
\label{approxsoltrav}
u_n(t)\approx \pm\,
 \tilde{v}_\alpha [\,  \sqrt{(\alpha -1)}\, (n-v_{\rm{s}} t -\xi_0 )  \, ],
\end{equation}
where we have set
$\tilde{v}_\alpha (\xi )= (1-\mu )^{\alpha/(1-\alpha)}\, v_\alpha (\sqrt{12(1-\mu )}\, \xi )$,
$v_{\rm{s}}=1+c\, (\alpha -1)$, $c=\tilde{c}/2$,
introduced an additional phase shift $\xi_0 \in \mathbb{R}$
and used the fact that $\epsilon = \sqrt{ \alpha -1 }$.
For any fixed value of $c$ and $\xi \in \mathbb{R}$, we have
$\lim_{\alpha\rightarrow 1^+}{\tilde{v}_\alpha (\xi )}=e^{2c + \frac{1}{2}-3\xi^2}\, $.
In this limit, the compacton (\ref{approxsoltrav}) converges towards the Gaussian
approximation (\ref{approxsol}), therefore it is also close to
the exact FPU solitary wave (\ref{exactvqqe}).

\section{\label{alt}Discussion}

We have obtained two generalized KdV equations (the log-KdV equation (\ref{logkdv}) and
the H-KdV equation (\ref{rac})) as formal asymptotic limits of Fermi-Pasta-Ulam lattices with
Hertzian-type potentials, when the nonlinearity exponent $\alpha >1$ goes to unity
and slowly varying profiles are considered. Using numerical computations, we have checked that
FPU solitary waves converge towards Gaussian solitary waves and compacton solutions to these KdV equations
when $\alpha \rightarrow 1^+$ and for near-sonic wave speeds.
In addition, we have illustrated numerically the formation of
stable solitary waves after a localized velocity perturbation in the Hertzian FPU system (\ref{nc}) and (\ref{vhertz})
when $\alpha \approx 1$, a limit in which the propagating pulse becomes nearly Gaussian.
The linearized log-KdV equation preserves the spectral stability of solitary waves, which
is lost when using other formal (Boussinesq-type) continuum models.
While our study does not yield a complete proof of the asymptotic behaviour of
exact FPU solitary waves when $\alpha \rightarrow 1^+$,
it provides nevertheless an asymptotic framework to explain classical formal compacton approximations
\cite{ap,neste2}, whose justification remained unclear up to now.

It would be interesting to examine the
dynamical properties of the log-KdV and H-KdV equations for different classes of initial conditions.
Relevant questions include local well-posedness (or ill-posedness), derivation of
a priori bounds, global well-posedness (or blow-up), scattering of some initial data,
and nonlinear stability of solitary waves. In our context, the study of the
nonlinear orbital stability \cite{bona}
or asymptotic stability \cite{pw} of solitary waves
rises new difficulties, linked with the lack of
smoothness of the energy functional (\ref{energy}).

In addition, if the well-posedness of the Cauchy problem for the log-KdV or H-KdV
equation could be established for appropriate initial data,
one could then study analytically or numerically
their connection with the FPU system (\ref{nc}) with homogeneous potentials (\ref{vhom}).
An open question is to check that well-prepared initial data evolve
(up to higher order terms and on long finite times) according to the log-KdV or H-KdV
equation (in the same spirit as the justification of the classical KdV equation
for FPU chains \cite{bp,kal,sw}). This problem may be extended to the Hertzian FPU system
(\ref{nc}) and (\ref{vhertz}), at least close to a solitary wave solution and
for a suitable topology (i.e. using a weighted norm
flattening perturbations behind the propagating wave \cite{pego,pw}).
The construction of appropriate numerical methods for
the time-integration of the log-KdV or H-KdV equations
is of course also fundamental in this context.
Another open problem concerns the dynamical
stability of the solitary wave solutions of the FPU lattice (\ref{ncrel}).
Our proof of the linear orbital stability
of solitary waves for the log-KdV equation
could be useful in this context,
following the lines of \cite{pego}
(using the linearized log-KdV equation instead of the linearized KdV equation).
These open questions will be explored in forthcoming works.

\vspace{2ex}

\noindent
{\it Acknowledgements:}
G.J. is grateful to
V. Acary, B. Brogliato, W. Craig and Y. Starosvetsky
for stimulating discussions on this topic.
Part of this work was carried out during a visit of G.J. to the
Department of Mathematics at McMaster University,
to which G.J. is grateful for hospitality.
G.J. acknowledges financial support from the French Embassy in Canada and
the Rh\^one-Alpes Complex Systems Institute (IXXI). D.P. acknowledges financial support from the NSERC.

\vspace{2ex}

\noindent
{\it Author contributions:}
G.J. introduced the time-dependent logarithmic and H\"olderian KdV equations and their derivations from the Hertzian chain model.
Numerical computations were performed by D.P. (figures 1, 3, 4) and G.J. (figure 5).
All authors contributed to the linear stability analysis of Gaussian solitary waves.
D.P. showed the convergence of the Ahnert-Pikovsky compacton towards a Gaussian for Hertz force exponents close to unity.
All authors contributed to the writing and editing of the manuscript.

\end{document}